\documentclass[useAMS,usenatbib]{mn2e}
\usepackage{graphicx}

\def\beg{\begin{eqnarray}}
\def\ende{\end{eqnarray}}
\def\gsim{\lower.4ex\hbox{$\;\buildrel >\over{\scriptstyle\sim}\;$}} 
\def\lsim{\lower.4ex\hbox{$\;\buildrel <\over{\scriptstyle\sim}\;$}}

\def\curl{{\rm curl}} 
\def\Om{{\it \Omega}}
\renewcommand{\vec}[1]{\mbox{\boldmath $#1$}}

\title[Radiation zone magnetohydrodynamics]
{ The effective magnetic Prandtl number  in  magnetized and differentially rotating stellar radiation zones}
\author[G. R\"udiger, M. Schultz, L.L. Kitchatinov]
{
G. R\"udiger$^1$\thanks{E-mail: GRuediger@aip.de},
M. Schultz$^1$, L.L. Kitchatinov$^2$ 
\\
$^1$Leibniz-Institut f\"ur Astrophysik Potsdam, An der Sternwarte 16, 14482 Potsdam, Germany\\
$^2$ Institute for Solar-Terrestrial Physics, P.O. Box 664033, Irkutsk, Russia 
}

\begin{document}

\date{Accepted . Received ; in original form }

\pagerange{\pageref{firstpage}--\pageref{lastpage}} \pubyear{2014}

\maketitle

\label{firstpage}

\begin{abstract}
With application to inner stellar radiative zones, a linear theory is used
to analyze the instability of a dipole-parity toroidal background field, in
the presence of density stratification, differential rotation, and
realistically small Prandtl numbers. The physical parameters are the normalized latitudinal shear $a$ and the normalized field amplitude $b$. Only the solutions for the wavelengths with the maximal growth rates are considered. If these scales are combined to the radial values of velocity, one finds that  the (very small) radial velocity only depends slightly on $a$ and $b$, so that it can be used as the free parameter of the eigenvalue system. 

The resulting instability-generated tensors of magnetic diffusivity and eddy viscosity are highly anisotropic. The eddy diffusivity in latitudinal direction exceeds the eddy diffusivity in radial direction by orders of magnitude. Its latitudinal profile shows a strong concentration toward the poles which is also true for the effective viscosity which has been calculated via the angular momentum transport of the  instability pattern.  The resulting effective magnetic Prandtl number reaches values of $O(10^2)$, so that the differential rotation decays much faster than the toroidal background field, which is {the} necessary condition to explain the observed slow rotation of the early red-giant and sub-giant cores by means of magnetic instabilities.
\end{abstract}

\section{Introduction}
Hydromagnetic dynamos can be understood as a magnetic instability driven by a flow pattern in fluid conductors. There are, however, strong restrictions on the characteristics of flows that can excite dynamos \citep{E46}, as well as on the geometry of the resulting magnetic fields \citep{C33}. The restrictions even exclude any dynamo activity for a number of flows, for example, differential rotation alone can never maintain a dynamo. 

An open question is whether magnetic instabilities are able to excite a sufficiently complicated motion that together with the background flow can generate magnetic fields. \citet{TP92} suggested that nonuniformly rotating disks can produce a dynamo when magnetorotational (MRI) and magnetic buoyancy instabilities are active. Later on, numerical simulations have shown that the MRI alone may be sufficient for the accretion disk dynamo. However, at least for low magnetic Prandtl numbers, it remains unclear whether the MRI-dynamo is physical, or just a numerical artifact.

Another possibility was discussed by Spruit (2002), who suggested that differential rotation and a magnetic kink-type instability \citep{T57,T73} can together drive a dynamo in stellar radiation zones. If real, such a dynamo would be very important for angular momentum transport in stars. It taps energy from differential rotation, thus reducing the rotational shear. Radial velocity fluctuations converting toroidal magnetic field into poloidal field are necessary for such a dynamo. This dynamo, therefore, would necessarily also mix chemicals in stellar interiors, which may have consequences for stellar evolution. 

However, simply the action of differential rotation and a magnetic instability
converting toroidal field into poloidal does not guarantee such a dynamo, which
has thus far not been proven to exist. Doubts are especially directed to the kink-type instability which, in contrast to the MRI, exists even without differential rotation and develops at the expense of the magnetic energy. Detailed estimations of the dynamo parameters are thus necessary to assess any dynamo-efficiency of the Tayler instability. 

A basic aspect of turbulent dynamos is the ability of correlated magnetic
(${\vec b}$) and velocity (${\vec u}$) fluctuations to produce a mean
electromotive force along the background magnetic field ${\vec B}$, and also
along the electric current $\vec J$, i.e.
\begin{equation}
 \langle {\vec u}\times{\vec b}\rangle = \alpha{\vec B}- \mu_0 \eta_{\rm T}{\vec J} .
 \label{1}
\end{equation}
We computed the $\alpha$ effect of the current-driven instability for rigid rotation and for a 3D magnetic geometry in a previous paper (R\"udiger et al. 2012). However, because a magnetic-instability-induced radiative-zone dynamo
requires a differential rotation as an energy source, it remains to find the instability and the resulting electromotive force in the presence of a (weak) differential rotation. We have shown earlier that the interaction of even
current-free toroidal fields and differential rotation leads to the
appearance of an own instability where the growth rates (for fixed magnetic
field amplitude) switch from the low diffusion frequency to the high rotation frequency. It would therefore not be surprising if
the current-driven instability under the presence of differential rotation exceeded the power of the instability with no rotation or with rigid rotation.
Indeed, for toroidal fields in cylindrical geometry one can show that for rigid rotation the instability is suppressed for too weak toroidal  magnetic field (Acheson 1978; Pitts \& Tayler 1985; Tataronis \& Mond 1987), while it is re-animated by differential rotation (see R\"udiger et al. 2013).
In the present paper we shall apply a linear theory to show that similar results
are obtained even in non-uniformly rotating spheres with northern and southern magnetic belts
of opposite direction.

The formation of red giants without internal angular momentum transport would lead to steep radial profiles of the internal rotation law. Ceillier et al. (2012) report the Kepler-like profile $\Om \propto r^{-1.6}$ for the low-mass red giant KIC 7341231. The {\em Kepler} data, however, speak another language. According to the asteroseismology results, the core rotates only slightly faster than the outer convection zone (Deheuvels et al. 2012, 2014). Eggenberger et al. (2012) argue that an artificial viscosity of $3 \times 10^4$ cm$^2$/s may explain the surprisingly flat  rotation profiles. The outward flux of angular momentum then produces the observed spin-down of the inner radiative core. The general condition for this mechanism to work is that {\em the rotation profile decays faster than the magnetic field}. The magnetic Prandtl number represents the ratio of the decay time of the magnetic field and the decay time of the differential rotation. This  basic requirement is thus fulfilled when the effective magnetic Prandtl number is much larger than unity. We have thus to  study whether and under what circumstances the magnetic Prandtl number becomes large. 

The model and the stability analysis of this paper are close to those of \citet{KR08} and will be discussed here only briefly. The main idea is the use of the very small but necessarily nonzero radial flow component as the free parameter of the eigenvalue system, which is assumed to be proportional to the product of the critical length scale and the growth rate of the unstable mode. The special choice of this parameter always becomes unimportant for the results, which can be understood as the ratio of second-order quantities as there are the energy ratios and/or the effective magnetic Prandtl number formed by the diffusivity coefficients.

The dominant component of the magnetic field inside a star is  assumed to be the toroidal one, which can easily be produced by differential rotation by induction from  a weak poloidal field. The background toroidal field of our model consists of two latitudinal belts of opposite polarities
\begin{equation}
 {\vec B} = r\sin\theta\sqrt{\mu_0\rho}\
 \Om_\mathrm{A}(\mu) {\vec e}_\phi 
 \label{2}
\end{equation} 
(see Spruit 1999) with $\Om_\mathrm{A}(\mu)$ as the Alfv\'en frequency of the toroidal field. Spherical coordinates are used with the axis of rotation as the polar axis; ${\vec e}_\phi$ is the azimuthal unit vector and $\mu= \cos\theta$.
Field belts with equatorial symmetry are described by functions $\Om_\mathrm{A}$ which are symmetric with respect to the equator. On the other hand, the simplest case with two belts of opposite signs in the two hemispheres is $\Om_\mathrm{A}= b\ \Om\ \mu$. The latitudinal profile of (\ref{2}) then peaks in mid-latitudes at $\theta=45^\circ$ and $\theta=135^\circ$. The quantity $b$ is referred in the following as the {\em amplitude} of $\hat\Om_\mathrm{A}=\Om_\mathrm{A}/\Om$. Within the solar tachocline toroidal fields of (say) 1 kG are represented by $b\simeq 0.01$.

The background flow is simply
\begin{equation}
 {\vec U} = r\sin\theta \ \Om \ {\vec e}_\phi
 \label{3}
\end{equation}
with $\Om$ as the equatorial rotation rate. For $\Om_\mathrm{A}=\Om={\rm const}$, i.e. $b=1$,  one finds $ {\vec U}={\vec V}_{\rm A}$ with ${\vec V}_{\rm A}$ the Alfv\'en velocity, which is stable according to the formulation of Chandrasekhar (1956) for ideal, perfectly-conducting fluids.

More realistic $\Om_\mathrm{A}$ and $\Om$ are  radius-dependent, but this dependence is not of basic importance for the analysis of barytropic instabilities. The reason is that the stratification of the radiative core is stable with positive 
\begin{equation}
 N^2 = \frac{g}{C_\mathrm{p}}\frac{\partial S}{\partial r} ,
 \label{4}
\end{equation}
where $S$ is the specific entropy and $C_\mathrm{p}$ the specific heat for constant pressure. 
We shall see that the latitudinal shear of the global rotation is much more important. Let the rotation law be given by
\begin{equation}
 \Om= \Om_0 (1 - a \ \cos^2\theta)
 \label{Om} 
\end{equation}
with $a$ as the shear parameter of the normalized differential rotation.  $\Om_0$ is the angular velocity of the equator, positive $a$ describes (`solar-type') equatorial acceleration. This is the typical definition for the tachocline theory: the outer convection zone above the radiative interior generates equatorial acceleration at its bottom which defines the shear at the top of the radiation zone.

\section{The equations}
In the upper solar radiative core\footnote{$N/\Om \simeq 15$ for red giants} the buoyancy frequency $N$ is large compared to $\Om$ ($N/\Om \simeq 400$). The radial scale of unstable disturbances is then short, and the dependence of the disturbances on the radius can be treated by the local approximation with $\mathrm{exp}(\mathrm{i}kr)$. A normalized wavelength of the perturbations is 
\begin{equation}
 \hat\lambda = \frac{N}{\Om k r} 
 \label{5}
\end{equation}
with $kr\gg 1$. The most unstable disturbances have $\hat\lambda < 1$ so that the radial scale of the disturbances, $\lambda = \pi/k$, does not exceed 1000 km (for solar values), which is certainly shorter than both the radial scales of the toroidal field or the angular velocity of rotation. For such short scales the local approximation in the radial coordinate is always applicable. The unstable modes in our analysis, however, remain global in their horizontal dimensions. 

The system of the linear stability theory includes the equation
\begin{eqnarray}
 \frac{\partial{\vec u}}{\partial t}
 + \left({\vec U}\cdot\nabla\right){\vec u}
 + \left({\vec u}\cdot\nabla\right){\vec U}
 + \frac{1}{{\mu_0\rho}}\left(\nabla\left({\vec B}\cdot{\vec b}\right)-\right.&&
 \nonumber \\
 - \left. \left({\vec B}\cdot\nabla\right){\vec b}
 - \left({\vec b}\cdot\nabla\right){\vec B}\right)
 =-\left(\frac{1}{\rho}\nabla P\right)' + \nu\Delta{\vec u} &&
 \label{6}
\end{eqnarray}
for the velocity disturbances, the equation
 \begin{equation}
 \frac{\partial{\vec b}}{\partial t} =
 \curl \left( {\vec U}\times{\vec b}
 + {\vec u}\times{\vec B} - \eta\nabla\times{\vec b}\right) 
 \label{7}
\end{equation}
for the magnetic disturbances, and
\begin{equation}
 \frac{\partial s}{\partial t} + {\vec U}\cdot \nabla s +
 {\vec u}\cdot\nabla S = \frac{C_\mathrm{p}\chi}{T}\Delta T'
 \label{8}
\end{equation}
for the entropy disturbances $s$.
Here $T+T'$ is the total fluctuating temperature. The equations (\ref{6})--(\ref{8}) were reformulated in terms of scalar potentials for toroidal and poloidal parts of the magnetic and velocity fields, i.e.
\begin{equation}
 \vec{u} =-\curl((rW)\vec{r}) -\curl\ \curl((rV)\vec{r})
 \label{9a}
\end{equation}
and
\begin{equation}
 \vec{b} = -\curl ((rB)\vec{r}) -\curl\ \curl((rA)\vec{r}).
 \label{9b}
\end{equation}
 \citep{C61}. The resulting system of five eigenvalue equations can be found elsewhere \citep{RKH13} and will not be reproduced here. The eigenvalue problem is solved numerically.

The equations include finite diffusion. Thermal diffusion is especially important because of its destabilizing effect (Acheson 1978). The radial perturbations  are suppressed   in radiation zones by buoyancy. The thermal diffusion smooths out the entropy disturbances, reducing the effect of stable stratification.  The diffusivities enter the normalized equations via the normalized quantities
\begin{equation}
 \epsilon_\chi = \frac{\chi N^2}{\Om_0^3 r^2},\ \ \ \ \ \ \
 \epsilon_\eta = \frac{\eta N^2}{\Om_0^3 r^2},\ \ \ \ \ \ \
 \epsilon_\nu = \frac{\nu N^2}{\Om_0^3 r^2}.
 \label{9}
\end{equation}
 The values of $\epsilon_\chi = 10^{-4}$, $\epsilon_\eta = 4\cdot 10^{-8}$, and $\epsilon_\nu = 2\cdot 10^{-10}$ characterize the upper part of the solar radiation zone. It is characteristic for stellar plasma that the magnetic Prandtl number
($5\cdot 10^{-3}$) is much larger than the ordinary Prandtl number 
($2\cdot 10^{-6}$). Both Prandtl numbers match the real situation in solar-type radiation zones. They are so small that no hope exists to apply direct numerical simulations to the problem. The smallness of the Prandtl numbers even forms a massive challenge for the numerical procedures of the {\em linear} theory. Note that the numerical values of (\ref{9}) for the red subgiants  differ from the solar-type values resulting in larger Prandtl numbers (R\"udiger et al. 2014).

The stability problem allows two types of equatorial symmetry of unstable eigenmodes. We use the notations S$m$ (symmetric mode with azimuthal wave number $m$) and A$m$ (antisymmetric mode) for the single modes. S$m$ modes have vector
fields $\vec{b}$ which are mirror-symmetric about the equatorial plane (symmetric ${b_r}, {b_\phi}$ and antisymmetric ${b_\theta}$) and mirror-antisymmetric flows
$\vec{u}$ (symmetric ${u_\theta}$ and antisymmetric ${u_r}, {u_\phi}$). Am modes
have antisymmetric $\vec{b}$ and symmetric $\vec{u}$. The instability which is 
discussed here has only been found for non-axisymmetric disturbances with
$m=\pm 1$.

The growth rates of modes with $m\pm 1$  are rather small for weak fields of both symmetry types. For weak fields, $0.01 < b < 1$, the growth rates are fitted very well by the parabolic law $\omega_{\rm gr} \simeq 0.1 b^2 \Om$. For strong fields, $b>1$, they are proportional to the field strength $\omega_{\rm gr}\simeq \Om_\mathrm{A}$, without depending on the rotation rate. There are obviously two different non-axisymmetric instabilities. The fast one is the standard Tayler instability of adiabatic fluids, while the slow one for finite heat-conductivity is of the double-diffusive type. The Tayler instability for adiabatic fluids exists for magnetic field configurations with equatorial symmetry only for $b > 1$, while this limit is reduced  if the dipolar field constellation of the present paper is applied. In this case a jump exists of the growth rates to higher values at $b = 1$. 

The azimuthal drift rates of the unstable $m=\pm 1$ modes also differ strongly for the two regimes, as they jump from resting to corotating just at $b=1$. 

The perturbations are considered as Fourier modes in time, azimuth and radius, of the form $\exp({\rm i}(kr+m\phi -\omega t))$. Only the highest-order terms in $kr$ are used (i.e. only the shortest waves) so that in radial direction the theory is a local one. The wave number $k$ enters the equations in the normalized form (\ref{5}) as a ratio of two large numbers.

 One can directly read from the equations that the radial flow component and the radial field component are much smaller than the horizontal components. If the fields are divergence-free then the immediate consequence is that for characteristic values $b_\phi/b_r\simeq kr\gg 1$ (and the same for $u_\phi/u_r$). In order to work with quantities of similar numerical values the notation $\hat u_r=(kr) u_r$ is used for estimates.

\begin{figure}
 \includegraphics[width=\columnwidth]{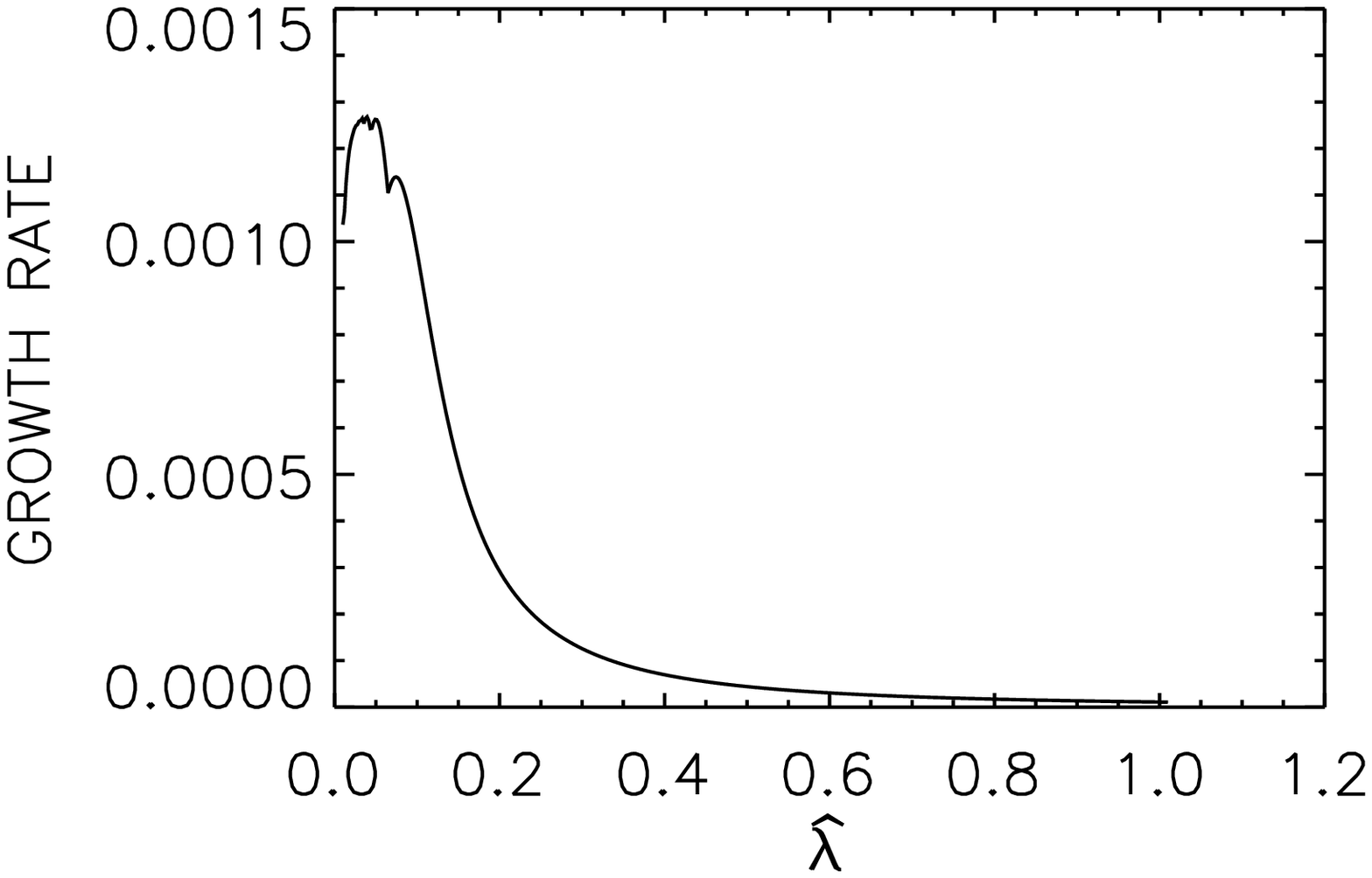}
 \includegraphics[width=\columnwidth]{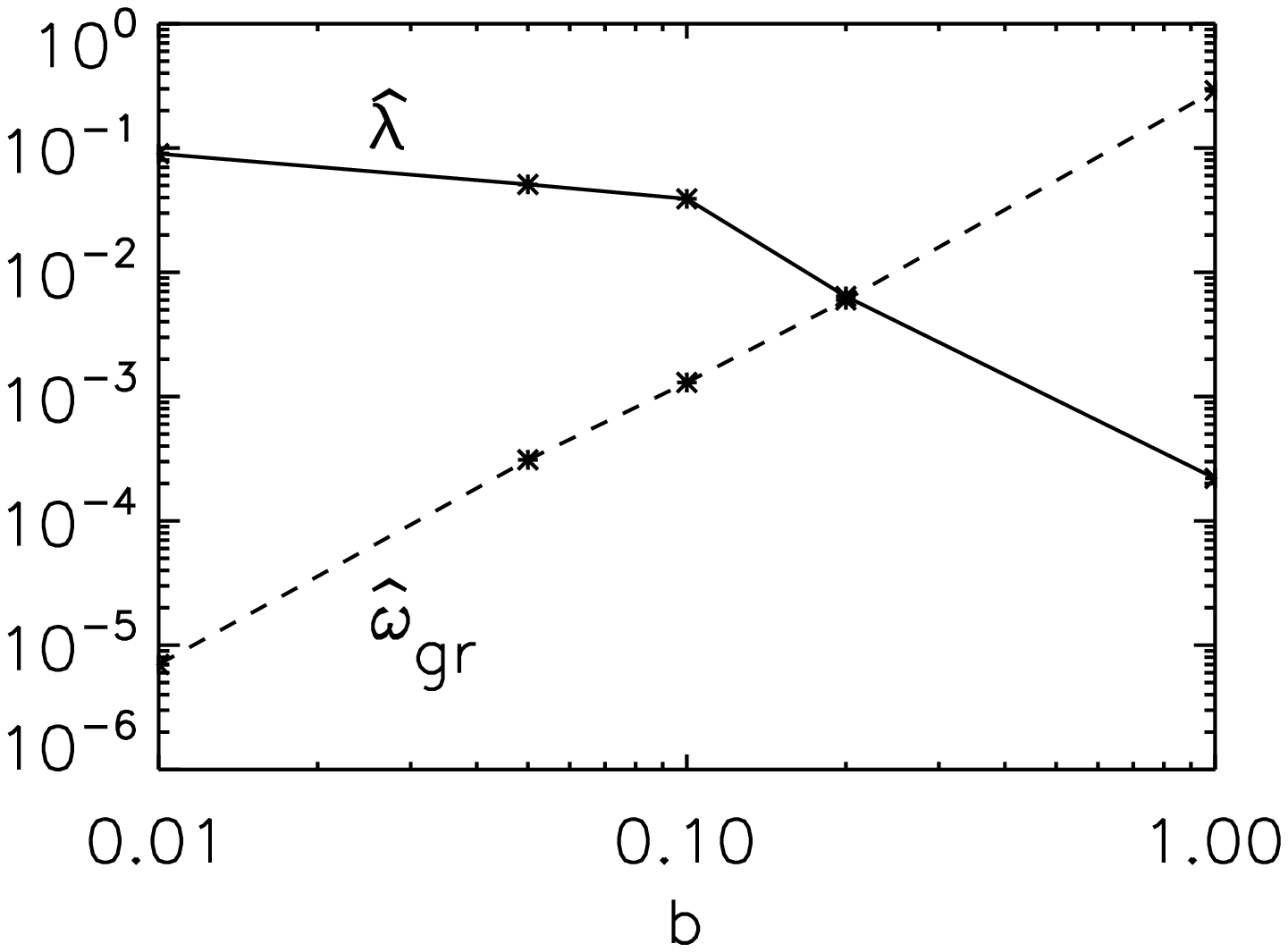}
\caption{Rigid rotation. Top: Growth rates $\hat \omega_{\rm gr}$ of the $m=\pm 1$ modes antisymmetric with respect to the equator for $b=0.1$ as a function of the normalized wavelength $\hat\lambda$. Bottom: The maximum growth rates and corresponding wavelengths for various magnetic field amplitudes $b$ taken from the Figs. \ref{gr1} and \ref{wl1}. $a=0$.}
 \label{solid}
\end{figure}

For rigid rotation the results are those given by R\"udiger et al. (2012). As an example Fig. \ref{solid} (top) gives the curves of the growth rates for $b=0.1$. One finds that for a characteristic wavelength $\hat\lambda$ the normalized growth rate has a maximum $\hat\omega_{\rm gr}$. The Fourier frequency $\hat\omega$ is a dimensionless quantity,  $\hat\omega=\omega/\Omega_0$.
For various $b$ these critical values of $\hat\lambda$ and $\hat\omega_{\rm gr}$ are collected in the lower panel of this figure. As a consequence of our normalizations they do not depend on the rotation rate $\Om$ or the buoyancy frequency $N$. One finds that for growing magnetic field amplitude $b$ the growth rate grows and the wavelength sinks. As a rough estimation the solutions of the equation system  can be summarized as $\hat\omega_{\rm gr}\simeq 0.1 b^2$  and $\hat\lambda\simeq 10^{-2}/\sqrt{b}$ for small $b$. Note that for $b\gsim0.1$ because of  $\hat\lambda\simeq 2\cdot 10^{-4}/{b^2}$ the product  $\hat\lambda\ \hat\omega_{\rm gr}$ hardly depends on the magnetic field amplitude.

\section{Differential rotation}
 We  shall now probe the influence of the shear on the excitation of the kink-type instability, which might be strong as the modes are non-axisymmetric. Generally, non-axisymmetric field components are suppressed by any form of differential rotation. We shall see, however,  that in our setup the differential rotation {\em destabilizes} the toroidal field belts.

\subsection{Growth rates}
\begin{figure*}
 \includegraphics[width=5.5cm]{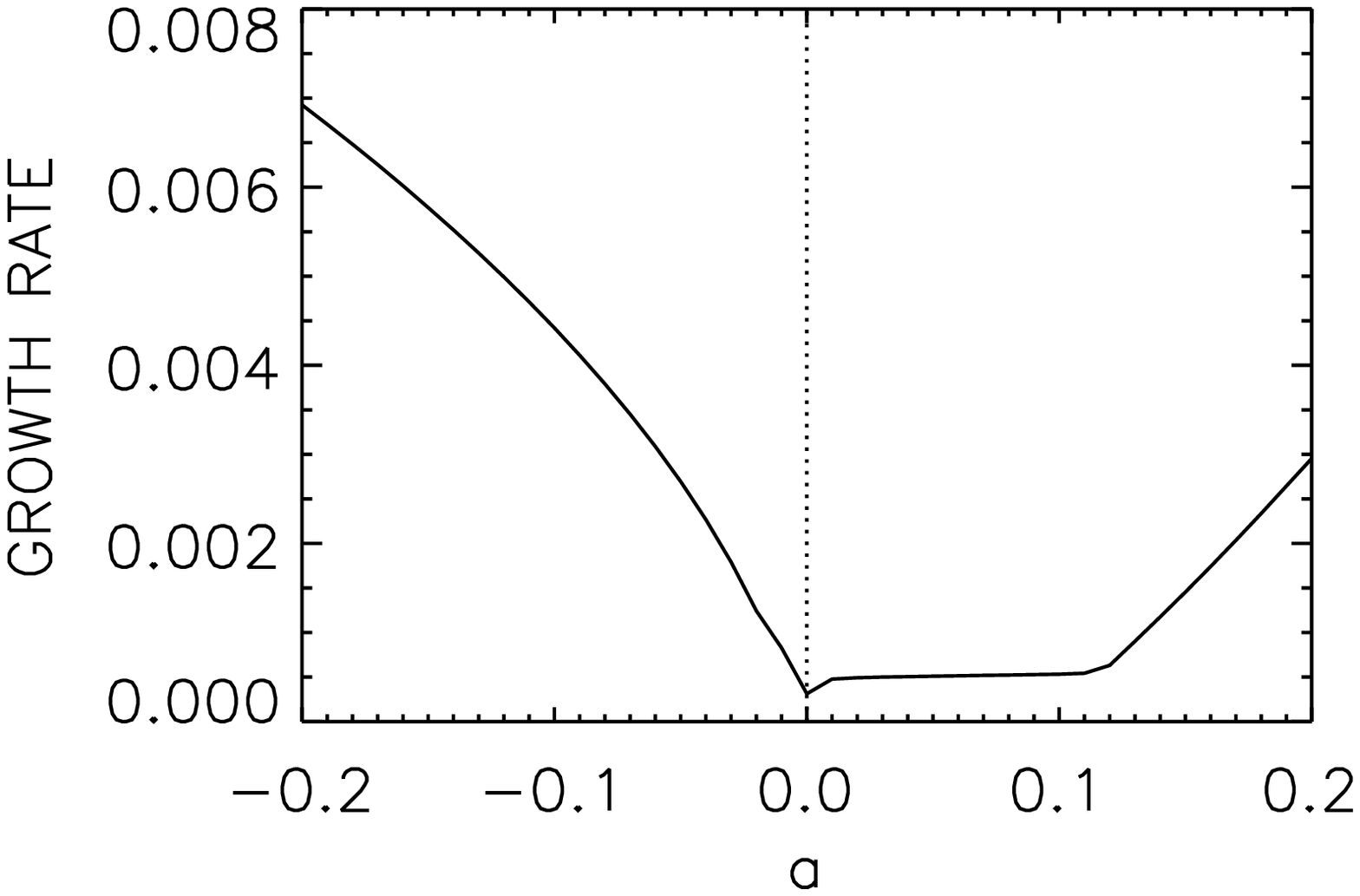}
 \includegraphics[width=5.5cm]{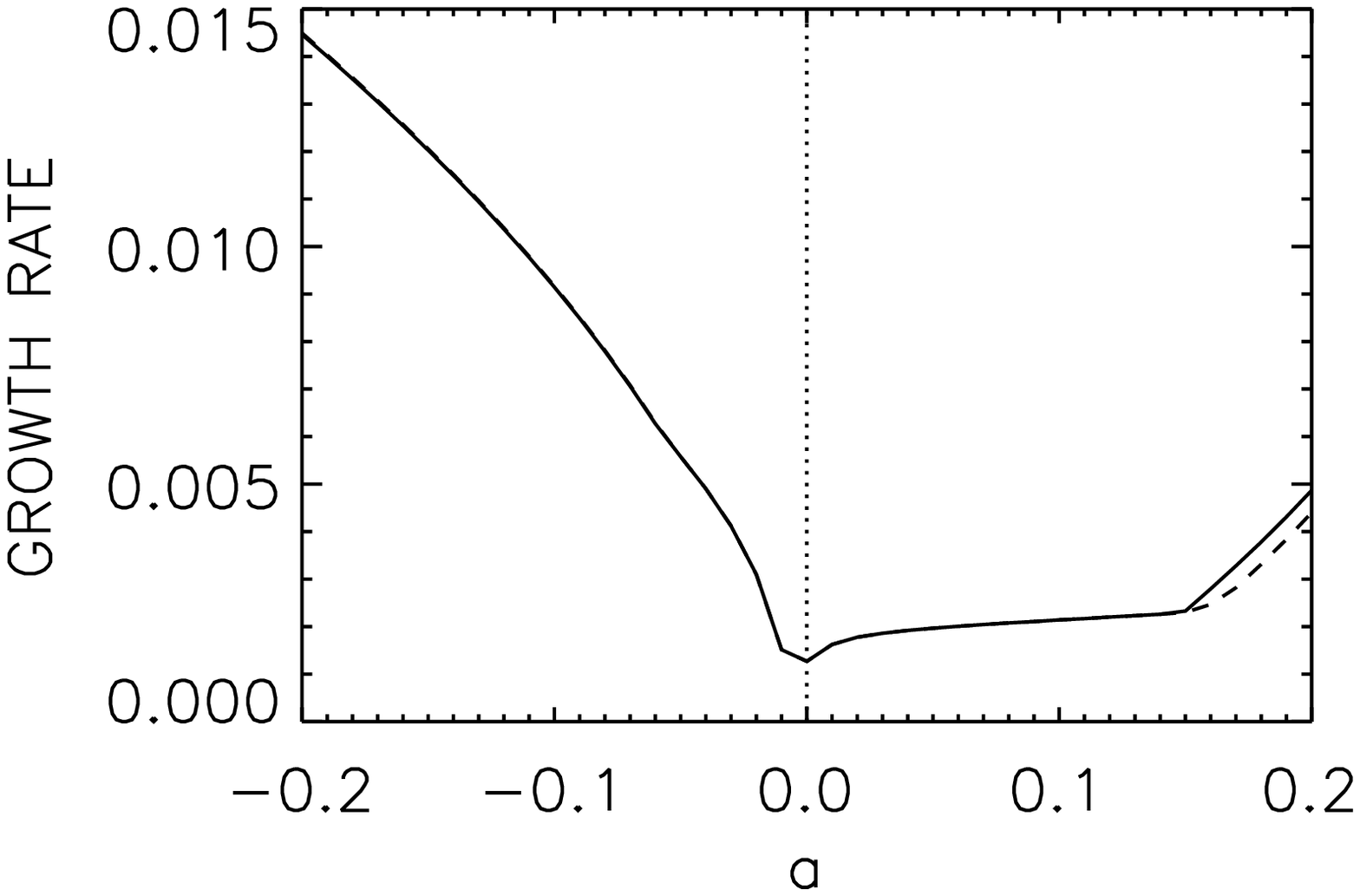}
 \includegraphics[width=5.5cm]{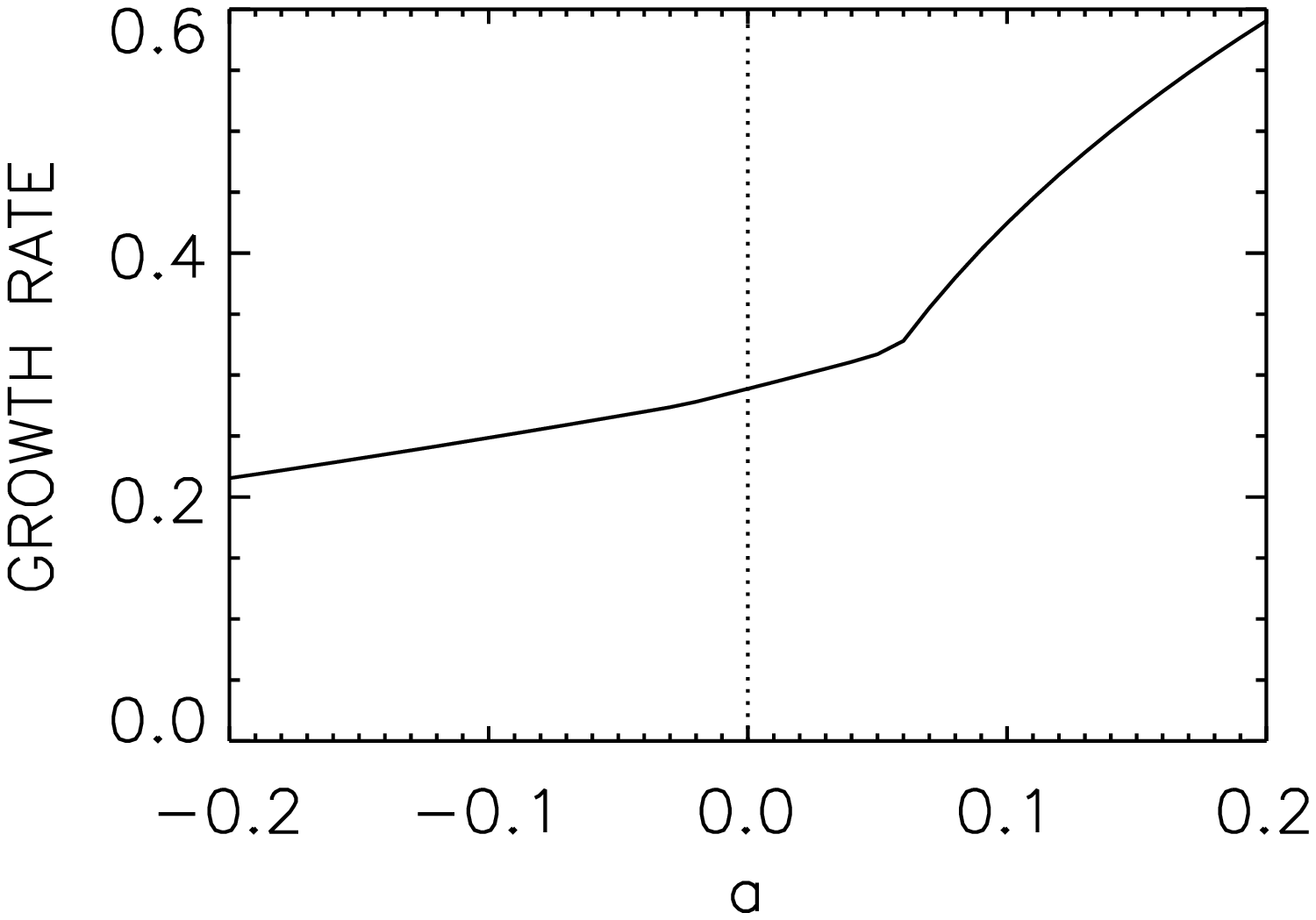}
\caption{Growth rates $\hat \omega_{\rm gr}$ -- normalized with the rotation rate -- of $m=\pm 1$ modes symmetric (solid) and antisymmetric (dashed) with respect to the equator, for rotation laws with finite shear. Left: $b=0.05$, middle: $b=0.1$, right: $b=1$. Positive shear always destabilizes the system. The background field with $b=0.01$ remains stable for positive shear (not shown).}
 \label{gr1}
\end{figure*}

The growth rates of the instability hardly depend on the symmetry type of the solutions with respect to the equator. Figure \ref{gr1} shows the growth rates for various shear parameters $a$ for the disturbances with equatorial symmetry (solid lines) and equatorial antisymmetry (dashed lines). The differences are strikingly small. More complex is the influence of the shear on the growth rates. The main result is that the shear supports the magnetic instability. 
Yet the growth rates do not simply depend on the value of $|a|$. There are also strong differences for weak and strong background fields.  For weak magnetic field the influence of positive and weak shear vanishes (Fig. \ref{gr1}, left panel). For the solar tachocline which may be characterized by $a\lsim 0.2$ and $b\simeq 0.01$ the stability results for rigid rotation thus remain valid also for inhomogeneous rotation laws. The growth rates for $b=0.05$  in  Fig. \ref{gr1} (left) are already positive but small for positive (weak) shear. One also finds that $b\simeq 0.01$ is {\em stable} for positive -- not too large -- shear.

For sufficiently strong shear the rotation energy is pumped into the instability modes and the growth rates grow for both signs of $a$. Figure \ref{gr1} also shows that negative shear   strongly destabilizes the toroidal field more than positive shear does.  Compared with the growth rates for rigid rotation a shear of only $a=-0.2$ increases the growth rate by one order of magnitude. This is true, however, only for $b<1$. For strong magnetic fields the negative shear slightly suppresses the instability while strong enough positive shear  destabilizes it (Fig. \ref{gr1}, right panel). The same phenomenon occurs in simulations of the kink-type instability for small magnetic Prandtl numbers in cylindrical geometry for strong fields.

 The interaction of shear and toroidal field modifies the relations between the growth rates and the magnetic field amplitudes. E.g., the growth rates for strong antisolar rotation laws ($a<0$) seem to indicate a linear dependence on $b$ rather than the standard $b^2$-relation.

\subsection{Drift rates}
The Tayler instability for $b\simeq 1$ is a  different phenomenon than the double diffusive instability for $b<1$. The drift rates differ strongly for weak and strong magnetic background fields, while the influence of $a$ is rather weak for weak fields. 
It has been found earlier that the modes with $m\pm 1$ are corotating with the star for weak fields ($b<1$), but they are stationary in the inertial frame for stronger background fields.  Note that the non-axisymmetric Tayler instability without rotation is also stationary in the laboratory system and that for large $b$ the influence of the rotation becomes weaker and weaker. It becomes thus understandable that for $b\geq 1$ the instability pattern becomes stationary in the inertial frame.

The influence of the shear can easily be described for weak field amplitudes. The drift in the corotating system is simply ${\dot\phi}/\Om_0\simeq -a/2$. It means that for a solar-type rotation law ($a>0$) the pattern no longer rotates with the equatorial rotation rate. The rotation rate of the poles is $\Om_0 (1-a)$, hence the pattern corotates with the mid-latitudes which is also true for negative $a$.
\subsection{Wavelengths}
\begin{figure}
 \includegraphics[width=\columnwidth]{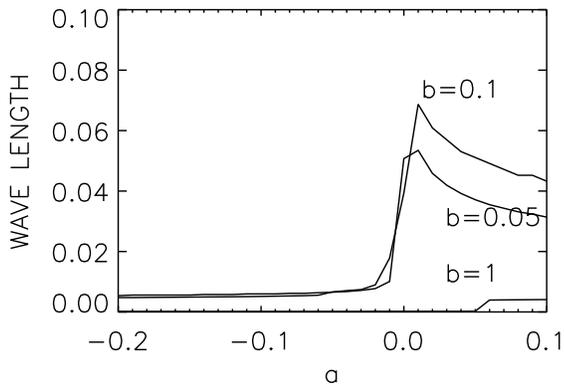}
\caption{Wavelengths $\hat\lambda$ for the $m=\pm 1$ modes antisymmetric with respect to the equator vs. rotational shear. The curves are marked with their values of the background field. Note the massive differences between the solutions with positive and negative shear. Strong fields $b\simeq 1$ only allow very short waves (see also Table 1).}
 \label{wl1}
\end{figure}

Also the  wavelengths of the instability pattern   depend on the applied shear {\em and} on the background-field amplitude (Fig. \ref{wl1}). Strong fields produce the smallest radial scales which hardly depend on the shear which is also known from the growth rates (see Fig. \ref{gr1}, right). For negative shear the wavelengths are nearly independent of $a$ and $b$. For weaker fields, however, and non-negative shear the results are much more complex. One finds that for weak fields the wavelengths are maximal for medium magnetic fields ($b\simeq 0.1$) but they are reduced for weaker and for stronger fields. For positive shear the normalized wavelengths exceed the wavelengths for negative shear by one order of magnitude. For weak fields this trend does not comply with the behavior of the growth rates as shown in Fig. \ref{gr1}.

 Of particular interest is the strong influence of very weak ($<  2$\%) positive and negative shear. For such values there is a steep and linear slope of the critical wavelengths with the numerical value of $a$ if the field is not too strong. Obviously, there are two different regimes for weak and for strong shear.

\subsection{Energies}\label{energy}

For rigid rotation we already know  that the kinetic energy ${\langle \vec{u}^2\rangle}$ for strong fields  is almost in equipartition with the magnetic energy ${\langle \vec{b}^2\rangle/\mu_0\rho}$ in contrast to the weak-field models with $b<1$ where the magnetic energy dominates  (R\"udiger et al.  2012). Here the ratio
\begin{equation}
\varepsilon_b=\frac{\langle \vec{b}^2\rangle}{\mu_0\rho \langle \vec{u}^2\rangle}
\label{ratio}
\end{equation}
of the two energies is calculated as averaged over latitude and longitude  under the influence of differential rotation.  
The radial field and flow values can be neglected in the calculation of (\ref{ratio}) as they are much smaller  than the horizontal components. It is, nevertheless,  {\em not} allowed to ignore them entirely in the equations as they must remain finite in order that the instability mechanism works.

As always for the solutions of a homogeneous system of linear
eigen-equations, all components are scaled with one and the same arbitrary factor. This free factor disappears if ratios of quantities such as (\ref{ratio}) are considered which are even in $\vec{u}$ and/or $\vec {b}$. 

\begin{figure}
\includegraphics[width=\columnwidth]{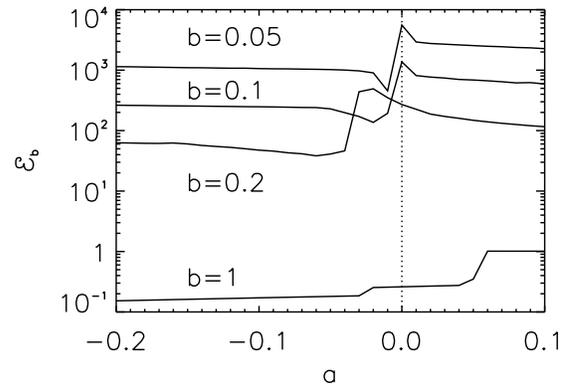}
 \caption{The ratio (\ref{ratio}) vs. the shear parameter $a$ between $b=0.05$ and $b=1$. Models with positive shear are more magnetically dominated than models with negative shear. Only for weak fields does the magnetic energy strongly exceed the kinetic energy.} 
 \label{energy3}
\end{figure}

Figure \ref{energy3} shows the ratio (\ref{ratio}) of the magnetic to the kinetic energy of the horizontal motions. For rigid rotation ($a=0$) the magnetic perturbations dominate the kinetic perturbations for the weakest given field ($b=0.05$). Generally, for $b\gsim 1$ the magnetic energy becomes almost equipartitioned with the kinetic energy.  A magnetic dominance of the instability only exists for the weak background fields with $b<1$ and the low growth rates and not for the strong fields with the high growth rates. Generally speaking the data of Fig. \ref{energy3} lead to the finding that for rigid rotation $\varepsilon_b\propto b^{-2}$ for $b\lsim 0.2$ and $\varepsilon_b\propto b^{-4}$ for $b> 0.2$.

The situation does hardly  change  for models with finite shear. The values of $\varepsilon_b$ are sligthly reduced for negative 
shear  and they are increased for positive shear  for all $b$.
If -- simplifying -- the Maxwell stress is considered to produce the eddy viscosity and the Reynolds stress is considered to produce the magnetic diffusivity then the effective (``turbluent'') magnetic Prandtl number 
\begin{equation}
{\rm Pm_{\rm T}}=\frac{\nu_{\rm T}}{\eta_{\rm T}}
\label{schmidt}
\end{equation}
should result as very large only  for small $b$. For strong fields with $b$ of order unity also the ratio (\ref{schmidt}) becomes O(1). In this case a possible differential rotation would decay with the same  time scale as the toroidal magnetic field. For large magnetic Prandtl numbers (\ref{schmidt}) the nonuniform rotation law decays much faster than the magnetic field.

In summary, for the given small microscopic Prandtl numbers the kinetic and magnetic patterns of the kink-type instability are anisotropic and magnetically
dominated for $b<1$ but they are isotropic and equipartitioned for background fields with $b\simeq 1$. Only positive shear produces  higher values for the energy \ref{ratio}) describing a dominance of the magnetic energy.
\subsection{Radial velocity}
\begin{figure}
 \includegraphics[width =\columnwidth]{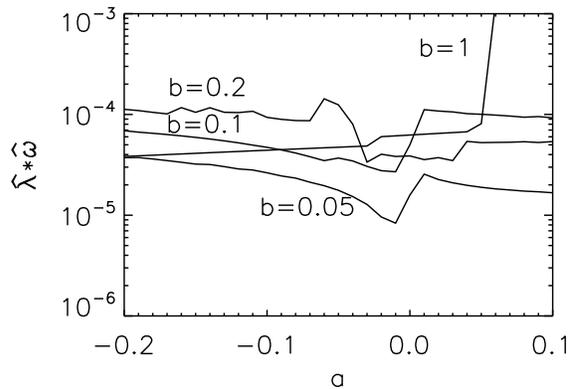}
\caption{The product ${\hat\lambda}{\hat\omega}_{\rm gr}$ for the $m=\pm 1$ modes antisymmetric with respect to the equator vs. the rotational shear. The influence of the shear remains weak for fields $b<1$.}
 \label{prod}
\end{figure}

It is clear that the calculation of the eddy diffusivity is complicated by the high degree of anisotropy in the flow and the field fluctuations, which are characteristic for instability in stratified media. It is also clear that a radial gradient of a toroidal field is dissipated by small-scale radial motions while a latitudinal gradient of the same field is dissipated by latitudinal small-scale motions. The eddy diffusivity in anisotropic turbulence is thus described by a tensor with different components in the radial and the horizontal directions. Here only that tensor component is considered which dissipates the {\em latitudinal} electric currents, i.e. 
\begin{equation}
 \eta_{rr}=\int_0^\infty\ \langle u_r(\vec{x},t)u_r(\vec{x},t-\tau)\rangle {\rm d} \tau\simeq u_{r}^2 \tau.
\label{etarr}
\end{equation}
The typical radial velocity perturbation $u_{r}$ can be estimated as follows. Write $u_{r}\simeq \lambda/\tau$ and use the inverse growth time $1/\omega_{\rm gr}$ as the timescale $\tau$. Hence,
\begin{equation}
u_{r}= \alpha\omega_{\rm gr} \lambda= \pi \alpha \frac{\Om}{N} \Om r\ (\hat\omega_{\rm gr} \hat\lambda),
\label{uu}
\end{equation}
where the free factor $\alpha$ serves as a tuning factor. It is the above mentioned  free unknown factor of the theory which always appears in linear and homogeneous equation systems. From Fig.~\ref{prod} one finds $\hat\omega_{\rm gr} \hat\lambda\lsim  10^{-4}$ so that with solar values $u_r\lsim 0.1$ mm/s. Equation (\ref{uu}) leads to the simple relation $\hat u_r=\alpha \pi \hat\omega_{\rm gr} \Om r$ which also gives the amplitudes of the horizontal components of the instability patterns ($u_{\rm rms}\simeq 1$ m/s for $b\simeq 0.1$). These amplitudes are  by the factor $\hat\omega_{\rm gr}$ smaller than the linear velocity of the global rotation.

From (\ref{etarr}) and (\ref{uu}) for small $b$  the estimate
\begin{equation}
\frac{\eta_{rr}}{\eta}\simeq 10^{-4} \alpha^2 (\frac{\Om}{N})^2\ b\ {\rm Rm},
\label{etarrr}
\end{equation}
follows with the magnetic Reynolds number ${\rm Rm}=\Om r^2/\eta $ which for the top of the solar radiative zone is of order $10^{12}$. It follows $\eta_{rr}/\eta\simeq 100 \alpha^2$. As this value has a reasonable order of magnitude there is obviously no need to use much smaller or much larger values for $\alpha$ than unity. Note. however, that  the horizontal components of the $\eta$-tensor should have much higher value becaus of the stron anisotropy existing in the system.

 The proxy ${\hat\lambda}{\hat\omega}_{\rm gr}$ for the radial velocity amplitude shows an almost uniform behavior concerning the shear for $b\gsim 0.1$ (Fig. \ref{prod}). One also finds that the shear hardly influences this product of the characteristic scales. Larger values only appear for strong shear and strong magnetic background field which is due to the increase of the characteristic wavelength $\hat\lambda$ shown in the lower right corner of Fig. \ref{wl1}.  It should thus be reasonable to use the relation (\ref{uu}) as the basis of the  calculations of effective magnetic
diffusivity and effective viscosity.  For the calculation of the effective magnetic Prandtl number (\ref{schmidt})  the tuning parameter $\alpha$ will not  influence the result.
 
\section{The electromotive force}\label{s4}
\begin{figure}
\includegraphics[width=\columnwidth]{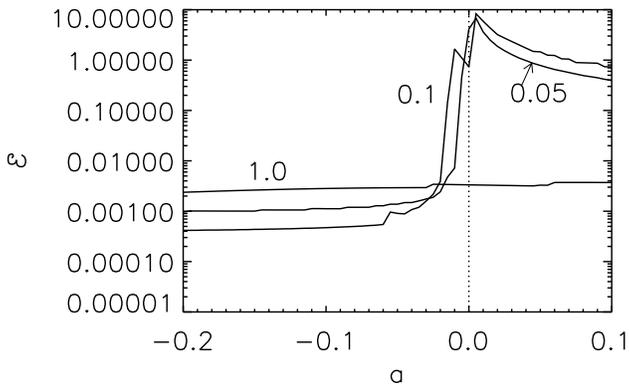}
\caption{The function $ \varepsilon$ in ots dependence on the values of $a$ and $b$. Note that $\varepsilon \simeq 0.03$ without any dependence on the rotation law.} 
 \label{epsilon2}
\end{figure}

According to Vainshtein \& Kichatinov (1983) the eddy diffusivity is based mainly on the Reynolds stress rather than the Maxwell stress. One may probe this idea with the linear instability theory by direct calculation of the electromotive force. After the definition 
\begin{eqnarray}
\langle \vec{u} \times \vec{b}\rangle_r = -\eta_{\theta\theta} {\rm curl}_r (B_\phi \vec{e}_\phi)
\label{ub}
\end{eqnarray}
of the eddy diffusivity $\eta_{\theta\theta}$ one finds 
\begin{equation}
 \eta_{\theta\theta}= - \frac{u_{\rm rms} v_{\rm rms} E_r}{b \Om}
\label{etat}
\end{equation}
with
\begin{equation}
E_r=\frac{\langle \vec{u} \times \vec{b}\rangle_r}{u_{\rm rms}b_{\rm rms} (3\cos^2\theta -1)}.
\label{Er}
\end{equation}
The Alfv\'en velocity $\vec{v}=\vec{b}/\sqrt{\mu_0\rho}$ of the magnetic perturbations $\vec{b}$ has been introduced. Because of the inclusion of the very small components $u_r$ and $b_r$  so that $\eta_{rr}\ll \eta_{\theta\theta}$. The $\eta$-tensor proves to be highly  anisotropic.
One finds
\begin{equation}
 \eta_{\theta\theta}= - \frac{E_r}{b} \varepsilon\frac{\langle {\hat u_r}^2\rangle}{\Om} \simeq -0.1 \alpha^2 \varepsilon E_r b^3\Om r^2,
\label{etar}
\end{equation}
where the ratio
\begin{equation}
\varepsilon= \frac{u_{\rm rms} v_{\rm rms}} {\langle {\hat u_r}^2\rangle}
\label{ratiou}
\end{equation}
 strongly depends an the values $a$ and $b$. As it should, however,  for strong fields ($b\simeq 1$) 
the $a$ dependence disappears (Fig. \ref{epsilon2}). On the other hand, for the considered small values of  $a$ and $b$ it results that $\varepsilon\simeq 0.01/b^2$.

\begin{figure}
\includegraphics[width=\columnwidth]{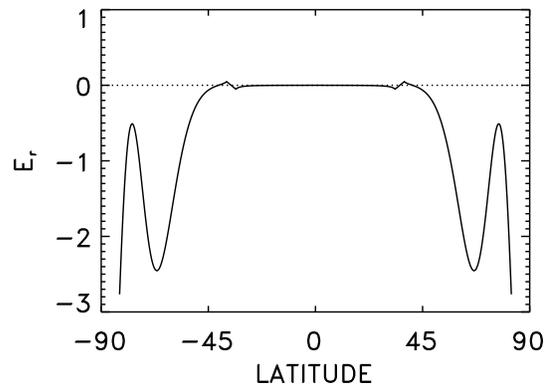}
\includegraphics[width=\columnwidth]{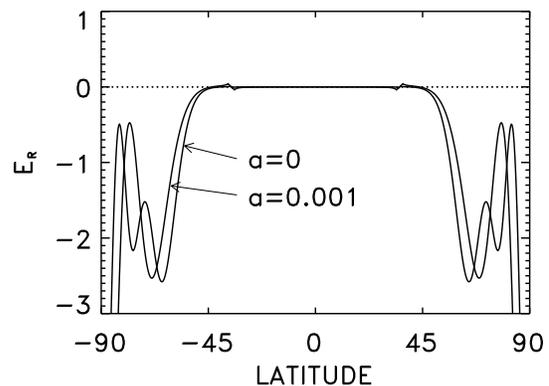}
\caption{The quantity $E_r$ for $b=0.05$ (top), $b=0.1$ (bottom, the value at the poles is $E_r\simeq-7$).The numerical value of $\overline{ E_r} $ (averaged over latitude) exhibits almost no dependence on $b$. There is also no strong influence of the weak shear on the values of $E_r$ (see the lower plot).} 
 \label{E1}
\end{figure}

Also  the dimensionless quantities $E$ can be computed with the linear and global code described above. The factor ${\Om r^2}$ carries the dimension of the diffusivities and can be estimated for the solar tachocline
 as $5 \cdot 10^{15}$ cm$^2$/s. 
All calculations lead to negative $E_r$. From Fig. \ref{E1} the $E_r$ does not vary strongly with $b$. To compress the data they are averaged over latitude. This procedure provides $\overline{E_r}\simeq -1$ for $b\ll 1$. Hence, the Eq. (\ref{etar}) provides magnetic-diffusivities which grow with $ b$ so that the diffusivity is amplified by the magnetic field, i.e.
\begin{equation}
\eta_{\theta\theta}\simeq 0.001 \alpha^2 b \Om r^2
\label{eta}
\end{equation}
One  finds $\eta_{\theta \theta}= 10^{12}\alpha^2 b$ cm$^2$/s for the eddy diffusivity in the upper domain of the solar radiative core.  For $b\simeq 0.01$ it results $\eta_{\theta \theta}= 10^{10}\alpha^2$ cm$^2$/s. As the component $\eta_{rr}$ of the diffusivity tensor is smaller by the factor $1/(kr)$ one finds $\eta_{rr}\simeq 10^7 \alpha^2$ cm$^2$/s for $b\simeq 0.01$. 

Note, however, that the relation (\ref{uu}) postulates the existence of a fluctuating radial flow  independent of the latitude which is certainly not quite realistic for the velocity pattern due to Tayler instability.

The described  antiquenching has serious consequences. The corresponding magnetic Reynolds number reads ${\rm Rm}_{\rm T}=\Om r^2/\eta_{\theta\theta}=1/(\eta_0 b)$ with $\eta_0=0.001 \alpha^2\overline{|E_r|}$. The amplitude of a toroidal field induced by the shear $a$  from a fossil poloidal field with the amplitude $b^{(0)}$ is $b/ b^{(0)}\simeq a {\rm Rm_T}$. Hence,
\begin{equation}
b\simeq \sqrt{\frac{a}{\eta_0}b^{(0)}}.
\label{rat}
\end{equation}
 For a shear of about 10\% and for $\alpha$ of order unity it yields $b\simeq 10 \sqrt{b^{(0)}}$. A poloidal 
fossil field of order 1 G in the solar tachocline region would thus lead to $b\simeq 0.01$ which is subcritical for $a>0$ (see the caption of Fig. \ref{gr1}).

\section{Angular momentum transport}\label{amt}
It is clear that the eddy viscosity cannot be estimated in the same way as the radial velocity (\ref{uu}), as the Maxwell stress may here be the dominant influence. The only possibility is thus to calculate the stress tensor (Reynolds stress plus Maxwell stress) to isolate the effective viscosity as the coefficient of the  shear. Again the expressions will be 
reformulated until they contain the radial velocity which finally will be expressed by means of the relation (\ref{uu}) including the tuning parameter $\alpha$. Such a procedure also leads to an effective viscosity as a function of $\alpha^2$.

The highly anisotropic velocity field which results from the magnetic instability is an almost horizontal one. It is known that rotating anisotropic turbulence fields behave in a non-Boussinesq way, in the sense that angular momentum is transported even for solid-body rotation. As this `$\Lambda$ effect' prevents rigid rotation from being a solution of the Reynolds equation, it has been used to explain the equatorial acceleration which can be observed at the solar surface and the surface of many solar-type stars (see R\"udiger 1989, with many references therein). It is shown in the following that the $\Lambda$ effect also occurs in the theory of magnetic instabilities, so that it is obvious that also in magnetized stellar radiative zones solid-body rotation cannot exist as a solution of the stationary equations. We shall show that the effect is only small for weak fields but its power grows for growing magnetic background fields.
 
In contrast to Spruit (2002) we focus on the latitudinal transport of angular momentum which is the natural choice if the rotation law has the structure (\ref{Om}) which is characteristic for radiative zones below a convection zone which drives a latitudinal $\Om$-gradient. The radial transport is weaker by orders of magnitudes although strong enough to produce the radially-uniform rotation of the solar interior.
The latitudinal angular momentum transport for latitudinal-dependent rotation rates is due to the combined action of Reynolds stress and Maxwell stress and can generally be written as
\begin{equation}
\langle u_\theta u_\phi \rangle-\frac{1}{\mu_0\rho}\langle b_\theta b_\phi\rangle = \nu_{\rm H}H \Om \cos\theta\sin^2\theta -\nu_{\rm H} \sin\theta \frac{{\rm d}\Om}{{\rm d}\theta},
\label{vis1}
\end{equation}
or with (\ref{Om}) as

\begin{equation}
\langle u_\theta u_\phi \rangle-\frac{1}{\mu_0\rho}\langle b_\theta b_\phi\rangle = \nu_{\rm H}\Om \cos\theta\sin^2\theta (H-2a).
\label{vis11}
\end{equation}
It consists of two parts. The first one also exists for rigid rotation 
 while the second one is of a diffusive nature and vanishes for vanishing shear. The $\Lambda$ effect always occurs for rotating 
density-stratified turbulence for sufficiently high rotation rate.
\begin{figure}
\includegraphics[width=\columnwidth]{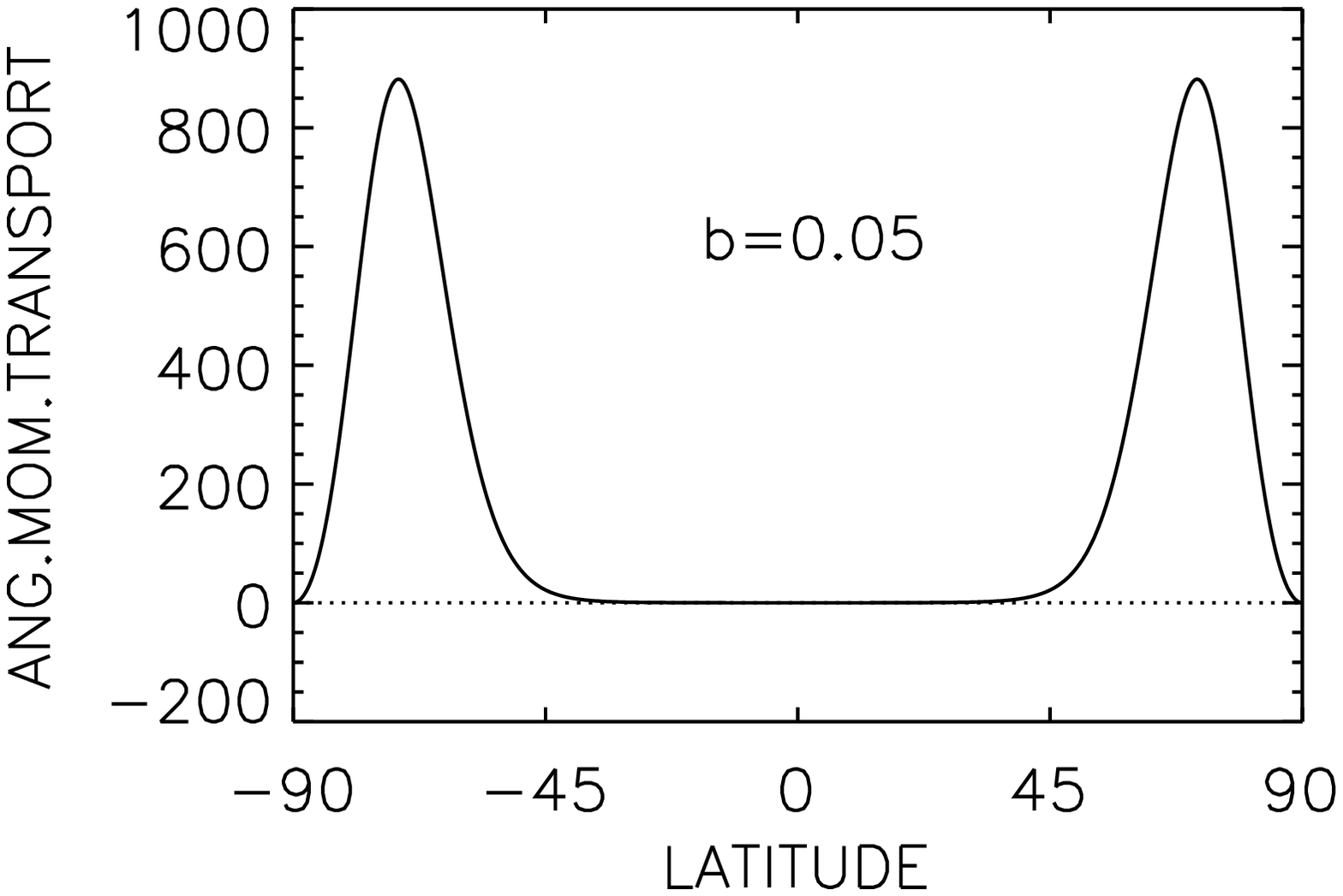}
\includegraphics[width=\columnwidth]{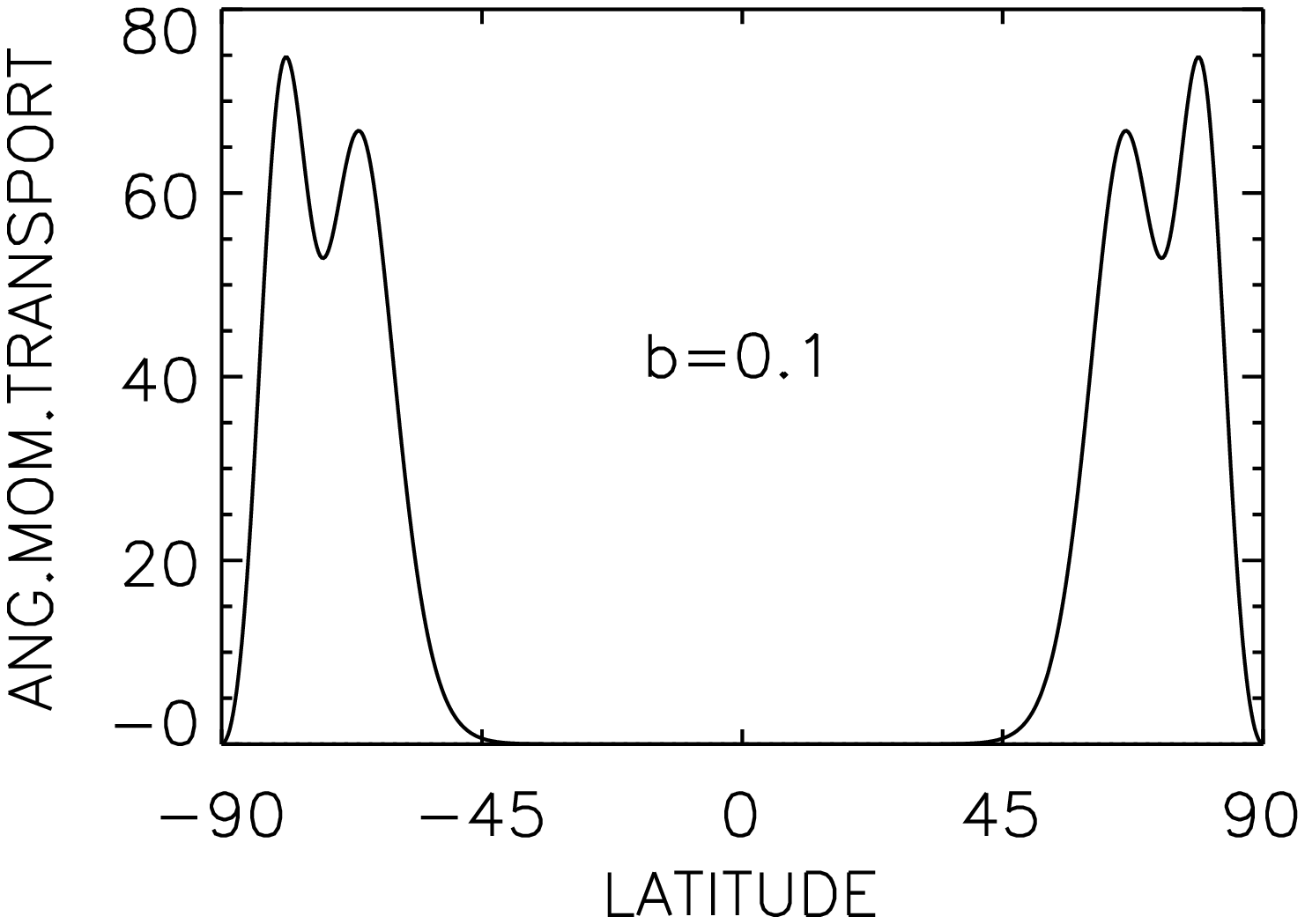}
\caption{The horizontal angular momentum transport  $q_\theta$ for $a=0$ and for various $b$. The effect vanishes rapidly for $b> 0.1$. The numerical value of $\overline{q_\theta}$ (averaged over latitude) depends strongly on $b$. The contribution of the Reynolds stress for the given examples is negligible in comparison to the Maxwell stress.} 
 \label{q1}
\end{figure}

It is possible to derive from (\ref{vis11}) the numerical values of the eddy viscosity and the $\Lambda$ effect in a similar way as the effective magnetic diffusivity has been derived in the previous Section. For $a=0$ we write 
\begin{equation}
 H\nu_{\rm H} = 2 \alpha^2 q_\theta (kr)^2 \frac{\langle u^2_r\rangle}{\Om}\simeq 0.2 \alpha^2 q_\theta b^4 \Om r^2, 
\label{vis2}
\end{equation}
which yields the product $H \nu_{\rm H}$. Here, the correlation function $q_\theta$ is defined by 
\begin{equation}
q_{\theta} = \frac{1}{2 \cos\theta\sin^2\theta}\frac{\langle u_\theta u_\phi \rangle-\langle b_\theta b_\phi\rangle/{\mu_0\rho}}{\langle {\hat u}^2_r\rangle},
\label{vis3}
\end{equation}
which is free of the arbitrary parameter $a$. Figures \ref{q1} show the numerical results for the expression (\ref{vis3}) for $a=0$ but for various values of the normalized magnetic field $b$. One finds that (\ref{vis2}) is positive for all $b$. In opposition to $E_r$ the $q_\theta$ possesses a strong inverse dependence on the amplitude of $b$. If again the data are averaged over latitude the relevant expression in (\ref{vis2})  then  $b^4\overline{q_\theta}$ runs with $ 0.01 b$. 

Obviously, the quantity $H$ is positive if the viscosity $\nu_{\rm H}$ is positive. This is true if, by (\ref{vis11}) for negative shear $a$ and positive $\nu_{\rm H} H$, the quantity $q_\theta$ is always positive, which can indeed be demonstrated.

\begin{figure}
\includegraphics[width=\columnwidth]{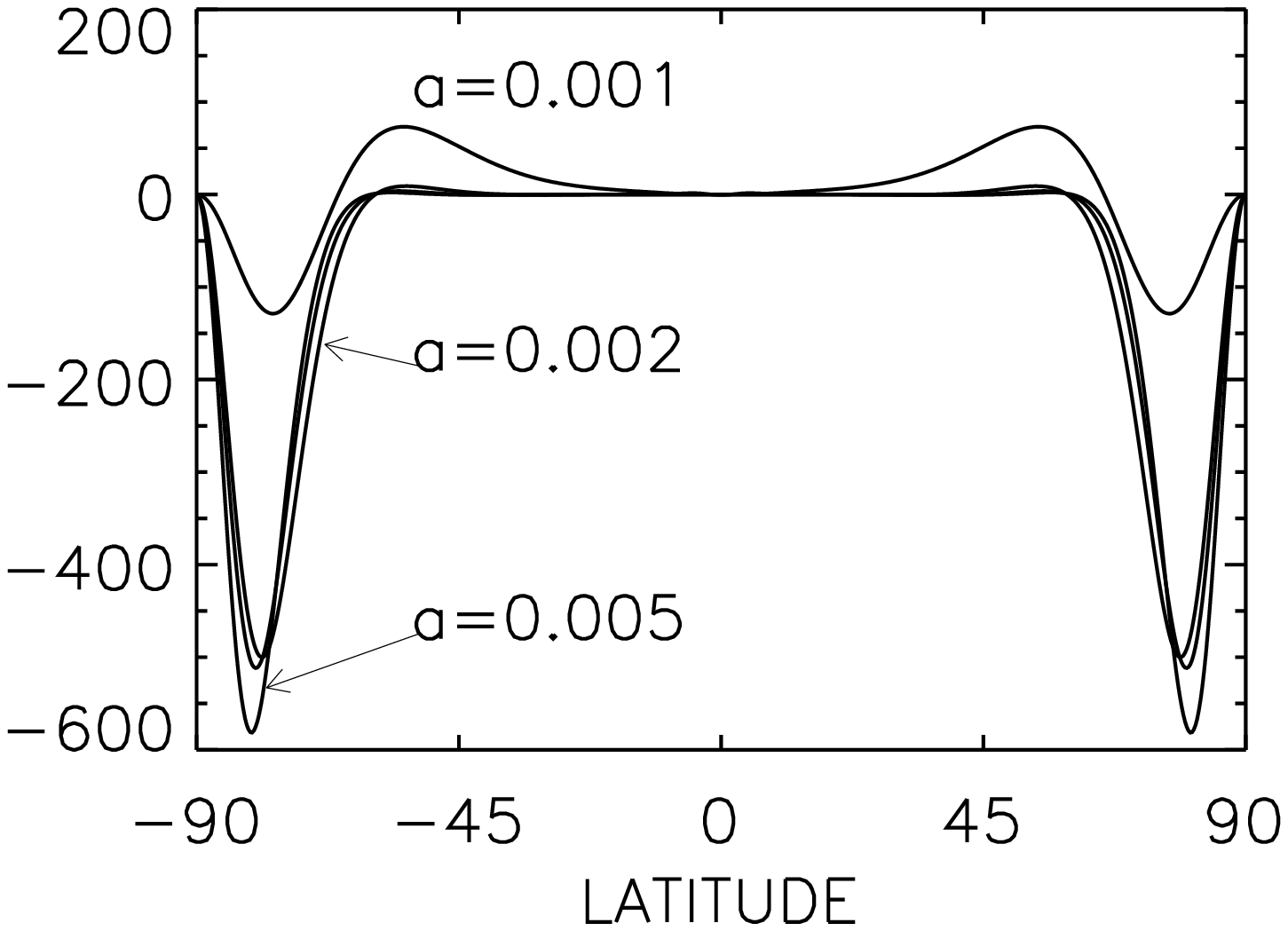}
\includegraphics[width=\columnwidth]{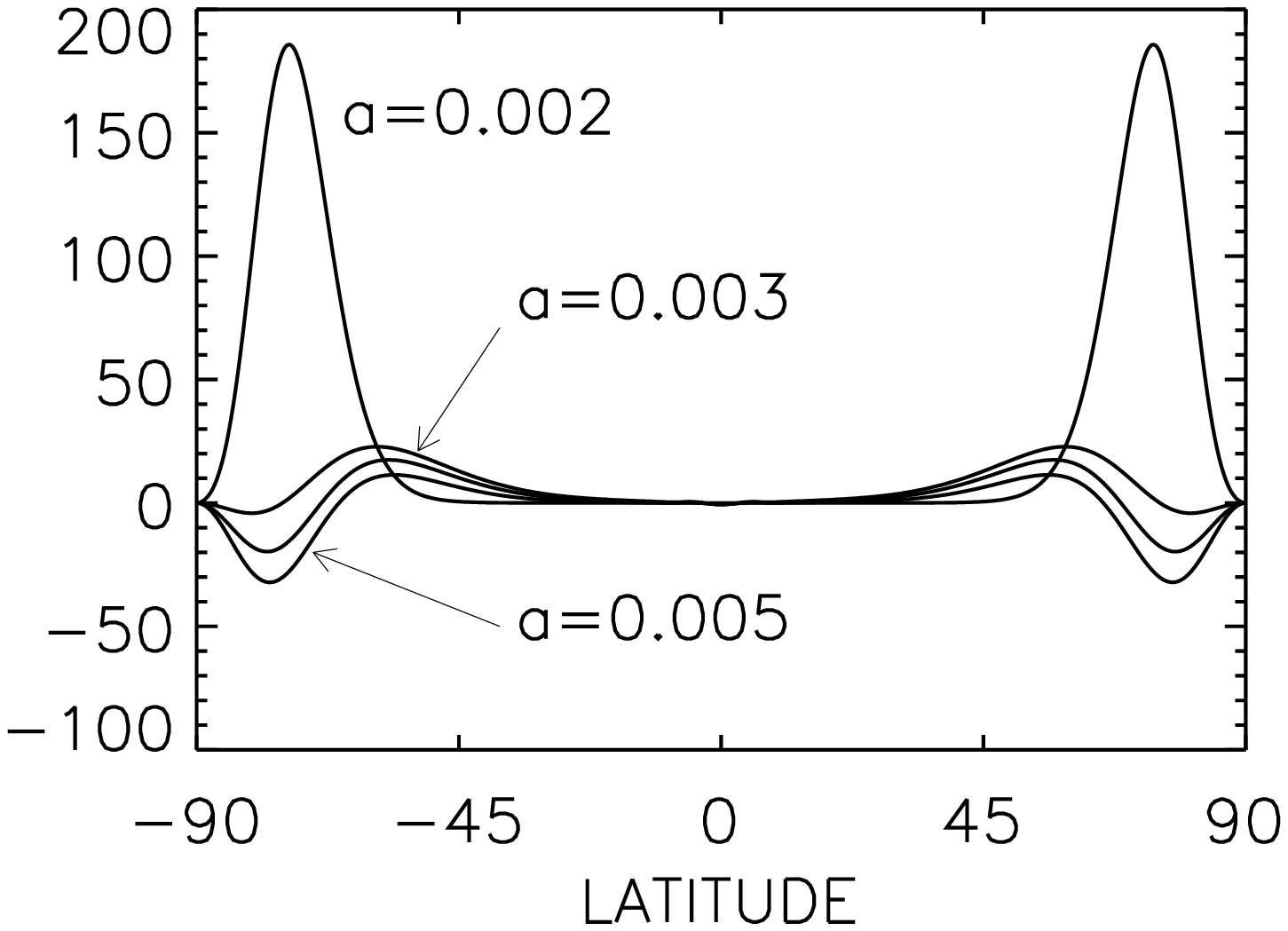}
\caption{The same as in Fig. \ref{q1} for various shear parameters. Top: $b=0.05$, bottom: $b=0.1$. The latitudinal average vanishes for $a_0\simeq 0.001$ (top) and $a_0\simeq 0.004$ (bottom).} 
 \label{q3}
\end{figure}

From (\ref{vis11}) the $\Lambda$-coefficient $H$ equals $2a$ if the correlation function $q_\theta$ provides very small values. If $H$ is known the viscosity $\nu_{\rm T}$ then directly follows from the result for $a=0$.
The dependence of the $q_\theta$ on the shear $a$ for the field with $b=0.1 $ is given in Fig. \ref{q3}. For finite and positive shear values $a_0$ the $q_\theta$ changes its sign. One reads from the plots that $H\simeq 0.002$ for $b=0.05$ and $H\simeq 0.008$ for $b=0.1$, hence $H\simeq 0.8 b^2$. From Eq. (\ref{vis2}) it follows
\begin{equation}
 \nu_{\rm H}\simeq 0.0025 \frac{\alpha^2}{b} \Om r^2.
\label{vis4}
\end{equation}
On the other hand, the $\Lambda$ effect prevents the formation of rigid rotation as a stationary solution of the Reynolds equation. For $H>0$ angular momentum is transported towards the equator and produces equatorial acceleration. It is also known that the resulting horizontal shear is approximated by the relation $a\simeq H/2=a_0$.  For the considered magnetized radiative stellar cores the resulting shear is less by two orders of magnitude than the observed shear in the solar convection zone.   Both the rotating convective envelope and the rotating magnetized radiative cores show equatorial acceleration -- the difference is mainly the magnitude.
\section{The effective magnetic Prandtl number}
For known $H$ the eddy viscosity $\nu_{\rm H}$ can be obtained from the RHS of (\ref{vis2}) by dividing by $H$. The combination of this expression and the result (\ref{etar}) for the magnetic diffusivity yields the effective magnetic Prandtl number 
\begin{equation}
 {\rm Pm_{\rm T}} = \frac{2}{H\varepsilon} \frac{b q_\theta}{|E_r|}\simeq \frac{2.5}{b^2},
\label{Pm}
\end{equation}
where the latter relation  valids\footnote{the detailed results for $b=0.15$ are here not shown} for $0.05 \lsim b \lsim 0.15$. The ratio no longer contains the  parameters $\alpha$, $r$ and $\Om$. As expected (because of the influence of the Maxwell stress) it will be much larger than unity if $b\ll1$. For small $b$ the resulting viscosity by far exceeds  the  eddy diffusivity. For larger magnetic fields, however, the magnetic energy is reduced in relation to the kinetic energy so that  the effective magnetic Prandtl number may approach unity.

 This result complies to that given in Fig. \ref{energy3}, which means that the magnetic energy in the instability pattern exceeds the kinetic energy only for weak fields with $b<1$. The instability pattern of fields with $b\gsim 1$ is {\em not} 
magnetically dominated.

\section{Discussion}
The instability of a large-scale toroidal magnetic field system with dipolar parity under the influence of a global rotation with latitudinal shear has been studied.  We are able to compute simultaneously the mean stress tensor and the mean electromotive force from one and the same model, which allows the determination of both the effective diffusion coefficients $\eta_{\theta\theta}$ and $\nu_{\rm H}$. The solution of such linear equation systems always
contains an unknown arbitrary amplitude factor, which we have fixed in the present paper by the relation $u_r=\alpha \lambda \omega_{\rm gr}$ for the radial flow of the instability, with $\lambda$ as the critical wavelength and $\omega_{\rm gr}$ as the maximal growth rate of the modes. The factor $\alpha$   plays  the role of the free parameter of the system.

Almost all of the numerical applications in this paper concern the upper layer of the solar radiative zone, also called the solar tachocline. The buoyancy frequency in this region exceeds the rotation rate by a factor of about 400, which leads to rather slow radial perturbations in comparison to the horizontal ones. If the Alfv\'en frequency of the toroidal field is taken in units of the rotation rate, i.e. $b\simeq\Om_{\rm A}/\Om$, then weak fields are described by $b\simeq 0.01$ and strong fields by $b\simeq 1$. For rigid rotation and for $\alpha\simeq 1$ the resulting radial flow for the instability with $b\gsim 0.5$ is of order 1 mm/s. The simplest estimate for the diffusion coefficient $\eta_{\rm T}\simeq \lambda^2\omega_{\rm gr}$ peaks for $b\simeq 0.1$ and reaches values of 250 in units of the microscopic diffusivity.
This value would lead to a decay of the toroidal magnetic background field after 0.5 Gyr. 
If the differential rotation shall decay within this time the condition $\rm Pm_{\rm T}>1$ must be fulfilled, otherwise the magnetic field decays faster than the differential rotation.

It is not possible for MHD flows to estimate the effective magnetic Prandtl number by simple scale relations, because Reynolds and Maxwell stresses contribute differently to the two diffusivities. To determine both the viscosity and the magnetic diffusivity 
from one and the same model one has to solve the full equation system and to form the correlations which define the two diffusivities. 
In ratios of expressions of second order such as the electromotive force and/or the Reynolds/Maxwell stress the free parameter $\alpha$ no longer influences the results.

The inspection of the correlations which provide eddy diffusivity and eddy viscosity reveals an extreme anisotropy of both diffusivity tensors. As the radial components of flow and field of the instability pattern are smaller by orders of magnitudes than the horizontal components, the correlations including $u_r$ and $b_r$ are also smaller by orders of magnitudes than the horizontal correlations. One must thus be careful with the definition of the magnetic Prandtl number. In this paper only the tensor components of the diffusivities are considered which dissipate both the latitudinal gradients of the background flow and field. Hence, the expressions (\ref{ub})  and  (\ref{vis1})  must be computed. The relation (\ref{Pm}) defines the magnetic Prandtl number as the ratio of the diffusion coefficients which result from the Eqs. (\ref{etar}) and (\ref{vis2}). We indeed find values up to $O(10^2)$ for the effective Prandtl number 
at $b\simeq 0.1$. 

 The magnetic diffusivity, however, grows monotonically for growing $b$. Strong fields with $b>1$ thus decay very quickly. For weaker background fields ($b\simeq 0.1$) the latitudinal shear of the global rotation decays 100 times faster than the field itself.
One might thus predict that already after (say) 1 Myr the radiative stellar cores rotate as a solid body. 

This, however, is not completely true. The calculations show that the  magnetic instability transports angular momentum equatorward even for solid-body rotation. This so-called $\Lambda$ effect avoids the formation of uniform rotation in the same way as it does in rotating solar and stellar convection zones where it is directly observable. Rigid rotation is no longer a solution of the stationary Reynolds equation with Lorentz force. As the density gradients and the horizontal anisotropies are weaker the $\Lambda$ effect for magnetized radiative zones, however, is much smaller than in rotating convection zones. Nevertheless, also weak differential rotation permanently produces toroidal fields from fossil poloidal fields, so that their existence during the existence of magnetic instabilities must be the rule rather than the exception.



\label{lastpage}

\end{document}